\documentclass{PoS}

\title{Review of low-energy interaction of strange and charm hadrons with nucleons and nuclei}

\ShortTitle{Review of low-energy interaction of strange and charm hadrons with nucleons and nuclei}

\author{\speaker{Laura Tolos}\\
        Instituto de Ciencias del Espacio (IEEC/CSIC), Campus Universitat 
Aut\`onoma de Barcelona, Facultat de Ci\`encies, Torre C5, E-08193 Bellaterra 
(Barcelona), Spain\\
Frankfurt Institute for Advanced Studies, Johann Wolfgang Goethe University, Ruth-Moufang-Str. 1,
60438 Frankfurt am Main, Germany\\
        E-mail: \email{tolos@ice.csic.es}}

\author{Carmen Garcia-Recio\\
        Departamento de F{\'\i}sica At\'omica, Molecular y Nuclear, and Instituto Carlos I de F{\'i}sica Te\'orica y Computacional,
Universidad de Granada, E-18071 Granada, Spain}
        
        \author{Raquel Molina\\
       Research Center for Nuclear Physics (RCNP),
Mihogaoka 10-1, Ibaraki 567-0047, Japan \\
        }
        
        \author{Juan Nieves\\
        Instituto de F{\'\i}sica Corpuscular (centro mixto CSIC-UV),
Institutos de Investigaci\'on de Paterna, Aptdo. 22085, 46071, Valencia, Spain\\
        }
        
        \author{Eulogio Oset\\
        Instituto de F{\'\i}sica Corpuscular (centro mixto CSIC-UV),
Institutos de Investigaci\'on de Paterna, Aptdo. 22085, 46071, Valencia, Spain\\
      }
        
                \author{Angels Ramos\\
        Departament d'Estructura i Constituents de la Mat\`eria,
Universitat de Barcelona,
Diagonal 647, 08028 Barcelona, Spain\\
     }
        
                        \author{Lorenzo Luis Salcedo\\
        Departamento de F{\'\i}sica At\'omica, Molecular y Nuclear, and Instituto Carlos I de F{\'i}sica Te\'orica y Computacional,
Universidad de Granada, E-18071 Granada, Spain\\
        }

\abstract{The properties of strange and charm mesons in nuclear matter and nuclei are reviewed. Different frameworks are presented and discussed paying a special attention to unitarized coupled-channel approaches. Possible experimental signatures of the in-medium properties of these mesons are also addressed, in particular in connection with the future FAIR facility at GSI.}

\FullConference{Xth Quark Confinement and the Hadron Spectrum,\\
		October 8-12, 2012\\
		TUM Campus Garching, Munich, Germany}

\begin{document}

\section{Introduction}
\label{intro}

Understanding the behavior of matter under extreme conditions of density and temperature has been a matter of debate over the last decades in order to gain insight not only into fundamental aspects of the strong interaction, such as the mass generation or deconfinement, but also into a variety of astrophysical phenomena, such as a the dynamical evolution of supernovae and the composition of neutron stars. The experimental programs of SIS/GSI, SPS/CERN, RHIC/BNL, the ongoing LHC/CERN project and the forthcoming PANDA and CBM experiments at FAIR/GSI are being developed to clarify the present situation.

A particular effort has been invested in understanding the properties of hadrons with strange and charm content in hot and dense matter. Since the pioneer work on kaon condensation in neutron stars \cite{kaplan}, the properties of mesons with strangeness in dense matter have been analyzed in connection to the study of exotic atoms as well as the analysis of heavy-ion collisions \cite{Fuchs:2005zg}. On the other hand, the study of the properties of open and hidden
charm mesons started more than 20 years ago in the context of relativistic nucleus-nucleus collisions in connection with charmonium suppression \cite{satz}  as a probe for the formation of quark-gluon plasma (QGP). The experimental program in hadronic physics at the future FAIR facility at GSI \cite{fair} will move from the light quark sector to the heavy one and will face new challenges where charm plays a dominant role. In particular, a large part of the PANDA physics program will be devoted to charmonium spectroscopy. Moreover, the CBM experiment will extend the GSI program for in-medium modification of hadrons in the light quark sector and provide the first insight into the charm-nucleus interaction.

In this paper we review the properties of the strange ($K$, $\bar K$ and $\bar K^*$)  and open-charm ($D$, $\bar D$ and $D^*$) mesons in dense matter. We address different approaches to obtain the in-medium properties of these mesons, paying a special attention to coupled-channels unitarized methods. Several experimental scenarios are also presented  so as to test the properties of these mesons in matter, such as heavy-ion collisions data on strange mesons, the transparency ratio of the $\gamma A \to K^+ K^{*-} A^\prime$ reaction and the formation of $D$-mesic nuclei.

\section{Strange mesons in matter}

\subsection{The properties of $\bar K$ mesons in a hot dense nuclear medium}

In the strange sector, the interaction of strange pseudoscalar mesons  with matter is a topic of high interest. Whereas the interaction of $\bar K N$ is repulsive at threshold, the phenomenology of antikaonic atoms shows that the $\bar K$ feels an attractive potential at low densities \cite{Friedman:2007zz}.

Antikaonic atoms, in which an electron is replaced by a negatively charged antikaon, can provide us with information about the interaction of an antikaon with nuclear matter. There have been some attempts to extract the antikaon-nucleus potential from best-fit analysis of antikaonic-atom data and some solutions, which use a phenomenological potential that includes additional non-linear density dependent terms, seem to be in agreement with very strongly attractive well depths of the order of -200 MeV at normal saturation density \cite{Friedman:2007zz}. However, some criticism has been raised due to the fact that antikaonic-atom data tests matter at the surface of the nucleus and, therefore, do not really provide a suitable constraint on the antikaon-nucleus potential at normal nuclear matter saturation density, $\rho_0=0.17 \ {\rm fm}^{-3}$.

Early works based on relativistic mean-field calculations using the meson-exchange picture or the chiral approach for the $\bar K N$ interaction while fitting the parameters to the $\bar KN$ scattering lengths \cite{Schaffner:1996kv} also obtained very deep potentials of a few hundreds of MeVs at saturation density. However, later approaches on unitarized theories in coupled channels based on the chiral approach \cite{Lutz,Ramos:1999ku} or on meson-exchange potentials \cite{Tolos01,Tolos02} obtain a potential much less attractive. 

In the unitarized coupled-channels models, the attraction is a consequence of the modified $s$-wave $\Lambda(1405)$ resonance in the medium due to Pauli blocking effects \cite{Koch} together with the self-consistent consideration of the $\bar K$ self-energy \cite{Lutz} and the inclusion of self-energies of the mesons and baryons in the intermediate states \cite{Ramos:1999ku}. Attraction of the order of -50 MeV at normal nuclear matter density  is obtained in unitarized theories in coupled channels based on chiral dynamics \cite{Ramos:1999ku} and meson-exchange models \cite{Tolos01,Tolos02}. Moreover, the knowledge of higher-partial waves beyond $s$-wave \cite{Tolos:2006ny,Lutz:2007bh,Tolos:2008di} becomes essential for relativistic heavy-ion experiments at beam energies below 2GeV per nucleon \cite{Fuchs:2005zg}. 

One of the latest unitarized approaches for calculating the antikaon self-energy in symmetric nuclear matter at finite
temperature is given in Refs.~\cite{Tolos:2008di,ewsr}. In this model, the antikaon self-energy and, hence, spectral function is obtained from the  in-medium antikaon-nucleon
interaction  in s- and p-waves within a chiral unitary approach.  The $s$-wave amplitude of the $\bar K N$ comes, at tree level, from the Weinberg-Tomozawa (WT) term of the chiral Lagrangian. Unitarization in coupled
channels is imposed on on-shell amplitudes ($T$) with a cutoff regularization. The $\Lambda(1405)$ resonance in the $I=0$ channel is generated dynamically and a satisfactory description of low-energy scattering observables is achieved. 
 
The in-medium solution of the $s$-wave amplitude accounts for Pauli-blocking effects, mean-field binding on the nucleons and hyperons via a $\sigma-\omega$ model, and the dressing of the pion and antikaon propagators. The self-energy is then obtained in a self-consistent manner summing the transition amplitude $T$ for the different isospins over the nucleon Fermi distribution at a given temperature, $n(\vec{q},T)$,  
\begin{eqnarray}
\Pi_{\bar K}(q_0,{\vec q},T)= \int \frac{d^3p}{(2\pi)^3}\, n(\vec{p},T) \,
[\, {T}_{\bar KN}^{(I=0)} (P_0,\vec{P},T) +
3 \, {T}_{\bar KN}^{(I=1)} (P_0,\vec{P},T)\, ], \ \  \ \ \ \label{eq:selfd}
\end{eqnarray}
where $P_0=q_0+E_N(\vec{p},T)$ and $\vec{P}=\vec{q}+\vec{p}$ are
the total energy and momentum of the antikaon-nucleon pair in the nuclear
matter rest frame, and ($q_0$, $\vec{q}\,$) and ($E_N$, $\vec{p}$\,) stand  for
the energy and momentum of the antikaon and nucleon, respectively, also in this
frame. The model includes, in addition, a $p$-wave contribution to the self-energy from hyperon-hole ($Yh$)
excitations, where $Y$ stands for $\Lambda$, $\Sigma$ and
$\Sigma^*$ components. The self-energy determines,
through the Dyson equation, the in-medium kaon propagator and the corresponding
antikaon spectral function:

\begin{eqnarray}
S_{\bar K}(q_0,{\vec q},T)= -\frac{1}{\pi}\frac{{\rm Im}\, \Pi_{\bar K}(q_0,\vec{q},T)}{\mid
q_0^2-\vec{q}\,^2-m_{\bar K}^2- \Pi_{\bar K}(q_0,\vec{q},T) \mid^2} \ .
\label{eq:spec}
\end{eqnarray}

\begin{figure}[htb]
\begin{center}
\includegraphics[width=0.7\textwidth]{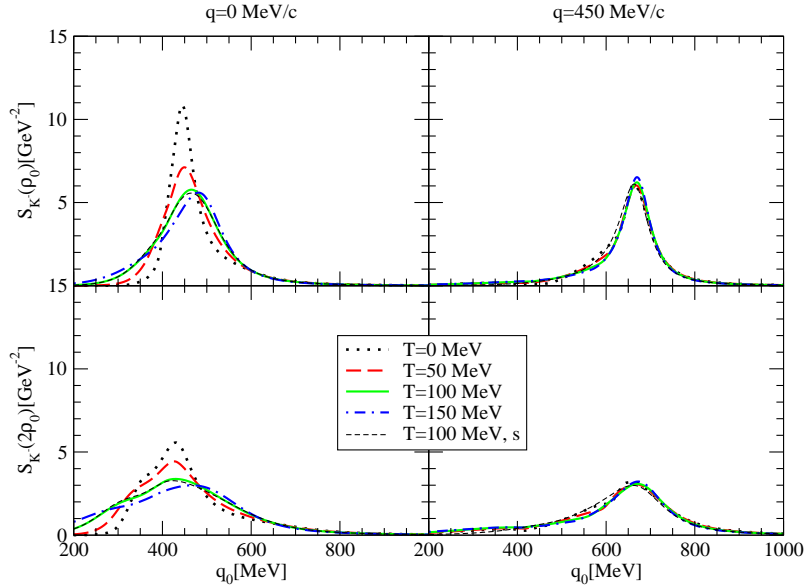}
\caption{ $\bar K$ spectral function for different densities, temperatures and momenta.}
 \label{fig1}
\end{center}
\end{figure}

The evolution of the  $\bar{K}$  spectral functions with density and
temperature is shown in Fig.~\ref{fig1}. The $\bar K$ spectral function
 shows a broad peak that results from a strong mixing between the
quasi-particle peak and the $\Lambda(1405)N^{-1}$ and
$Y(=\Lambda, \Sigma , \Sigma^*)N^{-1}$ excitations. 
These
$p$-wave $YN^{-1}$ subthreshold excitations affect mainly the properties of
the ${\bar K}$ at finite momentum, enhancing the low-energy tail of the spectral
function and providing a
repulsive contribution to the $\bar K$ potential that
partly compensates the attraction obtained from the stronger $s$-wave ${\bar K}
N$ interaction component. Temperature and density soften the $p$-wave
contributions to the spectral function at the quasi-particle energy. 

\subsubsection{Heavy-ion collision data for strange mesons}

From heavy-ion collisions there has been a lot of activity aiming at extracting the properties of  antikaons in a dense and hot environment. The antikaon production in nucleus-nucleus collisions at SIS
energies was studied some time ago using a Boltzmann-Uehling-Uhlenbeck (BUU) transport model with antikaons that were
dressed with the Juelich meson-exchange model \cite{Cassing:2003vz}.
Multiplicity ratios involving strange mesons coming from heavy-ion
collisions data were also analyzed \cite{Tolos:2003qj}. More recently,
a systematic study of the experimental results of KaoS collaboration
was performed together with a detailed comparison to transport model
calculations \cite{Forster:2007qk}. Several conclusions on the production
mechanisms for strangeness were achieved. However, the question that still remains is to which extend the
properties of $\bar K$ mesons are modified in matter. There is still no convincing
simultaneous description of all experimental data that involve antikaons in
matter. A recent report on
strangeness production close to threshold in proton-nucleus and heavy-ion
collisions  addresses the above issues \cite{Hartnack:2011cn}.

\begin{figure}[ht]
\begin{center}
\includegraphics[width=0.9\textwidth]{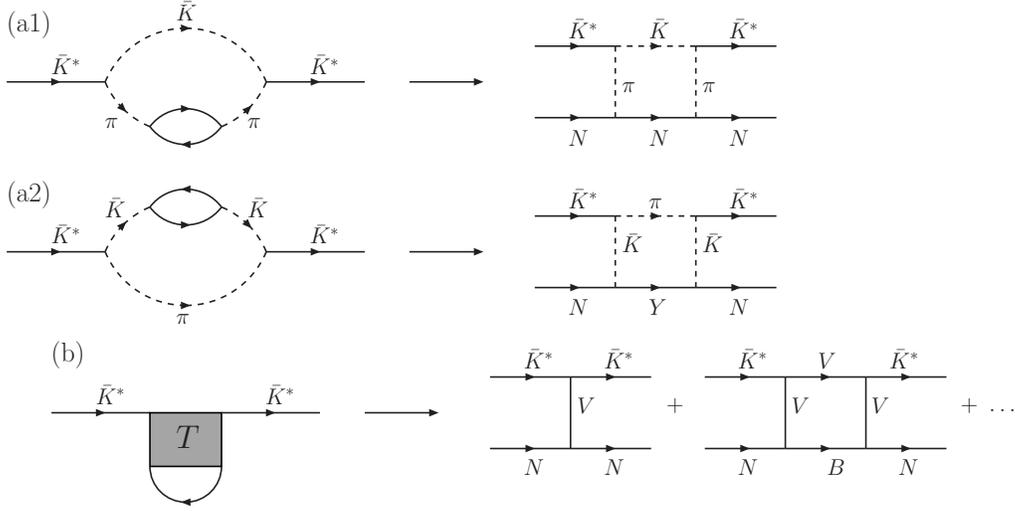}
\caption{Contributions to the $\bar K^*$ self-energy}
\label{fig:diag}
\end{center}
\end{figure}

\subsection{The $\bar K^*$ meson in nuclear matter}

With regard to vector mesons in the nuclear medium, only the non-strange vector mesons have been the main focus while 
very little discussion has been made about the properties of the strange ones,
$K^*$ and $\bar K^*$.  Most probably this is due to the fact that fstrange vector mesons can not be detected experimentally with dileptons.
Only recently the $\bar K^*N$ interaction in free space has been studied in
Ref.~\cite{GarciaRecio:2005hy} using SU(6) spin-flavour symmetry, and in
Refs.~\cite{Oset:2009vf, Khemchandani:2011et} within the hidden local gauge
formalism for the interaction of vector mesons with baryons of the octet and the
decuplet.  A recent review on results within the hidden gauge formalism for vector mesons 
can be found in Ref.~\cite{Oset:2012ap}.
 
The  $\bar K^*$  self-energy in symmetric nuclear matter is obtained  within the
hidden gauge formalism of Ref.~\cite{tolos10}. Two sources for the
modification of the $\bar K^*$ $s$-wave self-energy emerged in nuclear matter (see Fig.~\ref{fig:diag}): a) the
contribution associated to the decay mode $\bar K \pi$ modified by nuclear
medium effects on the $\pi$  and $\bar K$ mesons (see the (a1) and (a2) diagrams in Fig.~\ref{fig:diag}),  which accounts for
the $\bar K^* N \to \bar K N, \pi Y, \bar K \pi N, \pi \pi Y \dots$ processes, 
with $Y=\Lambda,\Sigma$, and b) the contribution associated to the interaction
of the $\bar K^*$ with the nucleons in the medium, which accounts for the direct
quasi-elastic process $\bar K^* N \to \bar K^* N$, as well as other absorption
channels involving vector mesons, $\bar K^* N\to \rho Y, \omega Y, \phi Y,
\dots$. In fact, this last term comes from a unitarized coupled-channel
process, similar to the $\bar K N$ case. Two resonances are generated
dynamically, $\Lambda(1783)$ and $\Sigma(1830)$, which can be identified with
the experimentally observed states $J^P=1/2^-$ $\Lambda(1800)$ and the
$J^P=1/2^-$ PDG state $\Sigma(1750)$, respectively \cite{Oset:2009vf}.
\begin{figure}[t]
\begin{center}
\includegraphics[width=0.5\textwidth,height=6cm]{spectral_ksn.eps}
\hfill
\includegraphics[width=0.4\textwidth,height=6cm]{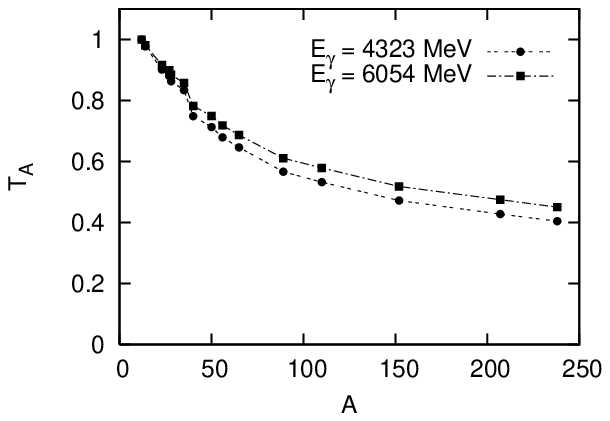}
\caption{ Left: $\bar K^*$ spectral function at $\vec{q}=0$ MeV/c for different densities. Right: Transparency ratio for  $\gamma A \to K^+ K^{*-} A'$}
\label{fig:spec-trans}
\end{center}
\end{figure}

The in-medium $\bar K^*$ self-energy results from the sum of both contributions,
$\Pi_{\bar K^*}=\Pi_{\bar{K}^*}^{{\rm (a)}}
+\Pi_{\bar{K}^*}^{{\rm (b)}}$,
 where $\Pi_{\bar{K}^*}^{{\rm (b)}}$ is obtained, similarly to the $\bar K$
meson in Eq.~(\ref{eq:selfd}), by integrating the $\bar K^* N$ transition
amplitude over the nucleon Fermi sea. The $\bar K^*$ spectral function is given in a similar way as in Eq.~(\ref{eq:spec}).

The $\bar K^*$ meson spectral function is displayed in the left panel of
Fig.~\ref{fig:spec-trans} as a function of the meson energy $q_0$, for zero
momentum and different densities up to 1.5 $\rho_0$. The dashed line refers to
the calculation in free space, where only the $\bar K \pi$ decay channel
contributes, while the other three lines correspond to the fully self-consistent
calculations,  which incorporate the process $\bar K^* \rightarrow \bar K
\pi$ in the medium, as well as the quasielastic $\bar K^* N \to \bar K^* N$ and
other $\bar K^* N\to V B$ processes.  The structures above the quasiparticle
peak correspond to the dynamically generated $\Lambda(1783) N^{-1}$ and
$\Sigma(1830) N^{-1}$ excitations. Density effects result in a dilution and
merging of those resonant-hole states, together with a general broadening of the
spectral function  due to the increase of collisional and absorption processes.
Although the real part of the optical potential is moderate, -50 MeV at
$\rho_0$, the interferences with the resonant-hole modes push the $\bar{K}^*$
quasiparticle peak to lower energies. However, transitions to
meson-baryon states with a pseudoscalar meson, which are included in the
$\bar K^*$ self-energy but are not incorporated explicitly in the unitarized
$\bar K^* N$ amplitude, would make the peak less prominent and difficult to
disentangle from the other excitations. In any case, what is clear from the
present approach, is that the spectral function spread of the $\bar K^*$
increases substantially in the medium, becoming at normal nuclear matter density
five times bigger than in free space.

\subsubsection{Transparency ratio for  $\gamma A \to K^+ K^{*-} A'$}

In order to test experimentally the $\bar K^*$ self-energy, one possible observable is the nuclear transparency ratio. The idea is to compare the cross
sections of the photoproduction reaction $\gamma A \to K^+ K^{*-} A'$ in
different nuclei, and trace them to the in medium $K^{*-}$ width.

The normalized nuclear transparency ratio is defined as
\begin{equation}
T_{A} = \frac{\tilde{T}_{A}}{\tilde{T}_{^{12}C}}, \hspace{0.5cm} {\rm with} \hspace{0.5cm} \tilde{T}_{A} = \frac{\sigma_{\gamma A \to K^+ ~K^{*-}~ A'}}{A \,\sigma_{\gamma N \to K^+ ~K^{*-}~N}} \ .
\end{equation}
 It describes the loss of flux of $K^{*-}$ mesons in the nucleus and is related to the absorptive part of the $K^{*-}$-nucleus optical potential and, thus, to the $K^{*-}$ width in the nuclear medium.  We evaluate the ratio between the nuclear cross sections in heavy nuclei and a light one ($^{12}$C), $T_A$, so that other nuclear effects not related to the absorption of the $K^{*-}$ cancel.

In the right panel of  Fig. \ref{fig:spec-trans} we show the results for
different nuclei. The transparency ratio has been plotted for two different
energies in the center of mass reference system, $\sqrt{s}=3$ GeV and $3.5$ GeV,
which are equivalent to energies of the photon in the lab frame of $4.3$ GeV and
$6$ GeV,  respectively. We observe a very strong attenuation of the $\bar{K}^*$
survival probability due to the decay  $\bar{K}^*\to
\bar{K}\pi$ or absorption channels 
$\bar{K}^*N\to \bar K N, \pi Y, \bar K \pi N, \pi \pi Y, \bar K^* N, \rho Y,
\omega Y, \phi Y, \dots$  with increasing nuclear-mass number $A$. This is due
to the larger path that the $\bar{K}^*$ has to follow before it leaves the
nucleus, having then more chances to decay or get absorbed.

\section{Charm mesons in matter}

 \subsection{Open charm in dense matter}

The medium modifications of mesons with charm, such as $D$ and $\bar D$ mesons, have been object of recent theoretical interest  due to the consequences for charmonium suppression, as observed at SPS energies by the NA50 collaboration \cite{NA501}. Furthermore, changes in the properties of the charm mesons will have a strong effect on the predicted open-charm enhancement in nucleus-nucleus collisions \cite{kostyuk,cassing}. 
 
A phenomenological estimate based on the quark-meson coupling (QMC) model \cite{Guichon:1987jp} predicts an attractive $D^+$-nucleus potential at normal nuclear matter density $\rho_0$ of the order of -140 MeV \cite{qmc}. The $D$-meson mass shift has also been studied using the QCD sum-rule (QSR) approach \cite{arata,weise,kaempfer}.  Due to the presence of a light quark in the $D$-meson, the mass modification of the $D$-meson has a large contribution from the light quark condensates. A mass shift of -50 MeV at $\rho_0$ for the $D$-meson has been suggested \cite{arata}. A second analysis, however, predicts only a splitting of $D^+$ and $D^-$ masses of 60 MeV at $\rho_0$ because the uncertainties to which the mass shift is subject at the level of the unknown $DN$ coupling to the sector of charm baryons and pions \cite{weise}.  Recent results on QSR rules for open charm mesons can be found in  \cite{kaempfer}.  On the other hand, the mass modification of the $D$-meson is also addressed using a chiral effective model in hot and dense matter \cite{dmeson}, where strong mass shifts were obtained.

With regard to approaches based on coupled-channel dynamics, unitarized coupled-channel methods have been applied in the meson-baryon sector with charm content \cite{Tolos:2004yg,Tolos:2005ft,Lutz:2003jw,Hofmann:2005sw,Mizutani:2006vq,Tolos:2007vh,Molina:2008nh,JimenezTejero:2009vq,JimenezTejero:2011fc}, partially motivated by the parallelism between the $\Lambda(1405)$ and the $\Lambda_c(2595)$. Other existing coupled-channel approaches are based on the J\"ulich meson-exchange model \cite{Haidenbauer:2007jq,Haidenbauer:2010ch} or on the hidden gauge formalism \cite{Wu:2010jy}.

More recently, an SU(6) $\times$ SU(2) spin-flavor symmetric model has been developed \cite{GarciaRecio:2008dp,Gamermann:2010zz}, similarly to the SU(6) approach in the light sector of Refs.~\cite{GarciaRecio:2005hy,Toki:2007ab}. This model implements heavy-quark spin symmetry (HQSS) \cite{Isgur:1989vq}, which is a proper QCD symmetry that appears when the quark masses, such as the charm mass, become larger than the typical confinement scale. The model generates dynamically resonances with negative parity in all the isospin, spin, strange and charm sectors  that one can form from an s-wave interaction between pseudoscalar and vector meson multiplets with $1/2^+$ and $3/2^+$ baryons \cite{Romanets:2012hm}. 

The HQSS predicts that, in QCD, all types of spin interactions involving heavy quarks vanish for infinitely massive quarks. Thus, HQSS connects vector and pseudoscalar mesons containing charm quarks.  Chiral symmetry fixes  the lowest order interaction between Goldstone bosons and other hadrons  in a model independent way; this is the Weinberg-Tomozawa  interaction. Then, it is very appealing to have a predictive model for four flavors including all basic hadrons (pseudoscalar and vector mesons, and $1/2^+$ and $3/2^+$ baryons) which reduces to the WT interaction in the sector where Goldstone bosons are involved and which incorporates HQSS in the sector where charm quarks participate. Indeed, this is a model assumption which is justified in view of the reasonable semiqualitative outcome of the SU(6) extension in the three-flavor sector \cite{Gamermann:2011mq} and on a formal plausibleness on how the SU(4) WT interaction in the charm pseudoscalar meson-baryon sector comes out in the vector-meson exchange picture.

Within this model, the self-energy and, hence, spectral function for $D$ and $D^*$ mesons are obtained self-consistently in a simultaneous manner, as it follows from HQSS, by taking, as bare interaction, the extended WT interaction previously described. We incorporate Pauli blocking effects and open charm meson self-energies in the intermediate propagators for the in-medium solution  \cite{tolos09}. Specifically, the $D$ and $D^*$ self-energies are obtained by summing the
transition amplitude over the Fermi sea of nucleons, as well as over the
different spin ($J=1/2$ for $DN$, and $J=1/2,3/2$ for $D^* N$) and isospin
($I=0,1$) channels, similarly to the $\bar K$ case.

\begin{figure}
\begin{center}
\includegraphics[width=0.6\textwidth]{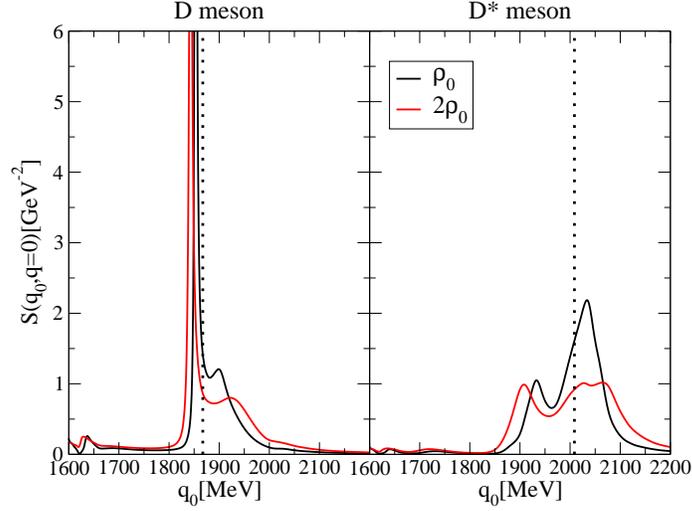}
\caption{The $D$ and $D^*$ spectral functions in dense nuclear matter at $\vec{q}=0$ MeV/c.}
\label{fig21}
\end{center}
\end{figure}

The $D$ and $D^*$ spectral functions at zero momentum are displayed in Fig.~\ref{fig21}. Those spectral functions show a rich spectrum of resonance-hole states. On one hand, the $D$ meson quasiparticle peak mixes strongly with $\Sigma_c(2823)N^{-1}$ and $\Sigma_c(2868)N^{-1}$ states. These $\Sigma_c$ resonances are predictions of our model with no experimental confirmation yet. On the other hand, the $\Lambda_c(2595)N^{-1}$ is clearly visible in the low-energy tail.  With regard to the $D^*$ meson, the $D^*$ spectral function incorporates the $J=3/2$ resonances, and the quasiparticle peak fully mixes with the  $\Sigma_c(2902)N^{-1}$ and $\Lambda_c(2941)N^{-1}$ states. The  $\Sigma_c(2902)$ can be identified with the experimental $\Sigma_c(2800)$ \cite{Nakamura:2010zzi} while $\Lambda_c(2941)$ has no experimental correspondance yet.  For both mesons, the $Y_cN^{-1}$ modes tend to smear out and the spectral functions broaden with increasing phase space, as seen before in the ${\rm SU(4)}$ model \cite{Mizutani:2006vq}. 

\begin{figure}
\begin{center}
\includegraphics[width=0.6\textwidth,height=0.5\textwidth]{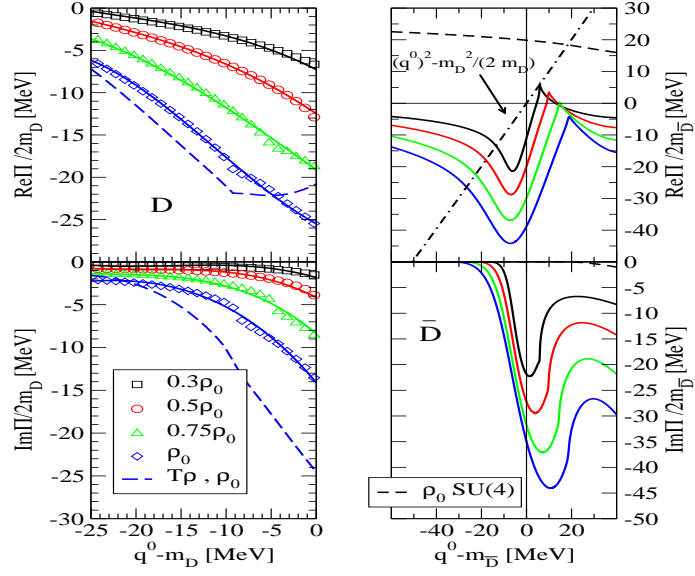}
\caption{ The $D$ and $\bar D$ optical potential at $\vec{q}=0$ MeV/c for different densities }
\label{fig22}
\end{center}
\end{figure}

The optical potential for $D$ and $\bar D$ meson is shown in  Fig.~\ref{fig22}. Defined as
\begin{equation}
  V(r,E) = \frac{
  \Pi(q^0=m+E,\vec{q}=0,~\rho(r))}{2 m},
\label{eq:UdepE}
\end{equation}
where $E=q^0-m$ is the $D$ or $\bar D$ energy excluding its mass, and $\Pi$ the meson self-energy, we observe in both cases a strong energy dependence close to the open-charm meson mass. 
In the case of the $D$ meson, this is due to the mixing of the quasiparticle peak with the  $\Sigma_c(2823)N^{-1}$ and $\Sigma_c(2868)N^{-1}$ states. For the $\bar D$ meson, the presence of a bound state at 2805 MeV \cite{Gamermann:2010zz}, almost at $\bar D N$ threshold, makes the potential also strongly energy dependent. This is in contrast to the SU(4) model (see Ref.~\cite{carmen10}).

\subsubsection{D mesons in nuclei}


In this section we present a possible experimental scenario for the detection of the changes in matter of the properties of open charm mesons. In particular, we would like to analyze the possible formation of $D$ mesic nuclei. 

The QMC model predicted $D$ and $\bar D$-meson bound states in $^{208}$Pb  relying upon an attractive  $D$ and $\bar D$ -meson potential in the nuclear medium  \cite{qmc}. The experimental observation of those bound states, though, might be problematic since, even if there are bound states, their widths could be very large compared to the separation of the levels. This is indeed the case for the potential derived from a SU(4) $t$-vector meson exchange model for $D$-mesons \cite{Tolos:2007vh}.

\begin{figure}[t]
\begin{center}
\includegraphics[width=0.4\textwidth,angle=-90]{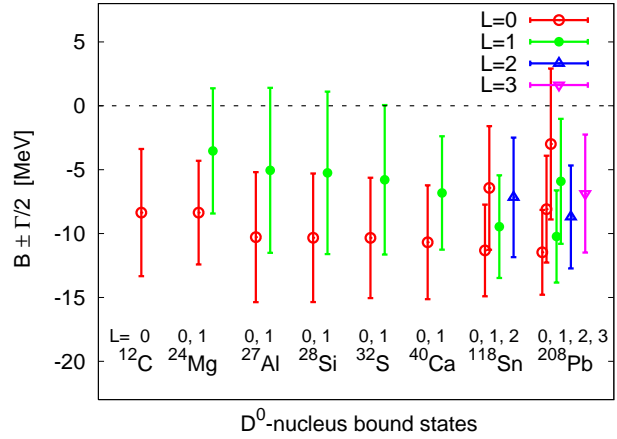}
\caption{$D^0$-nucleus bound states. \label{fig3}}
\end{center}
\end{figure}

We solve the Schr\"odinger equation in the local density approximation so as to analyze the formation of bound states with charm mesons in nucleus. In particular we study $D$ and $D^0$ mesons in nucleus. We observe that the $D^0$-nucleus states are weakly bound (see Fig.~\ref{fig3}), in contrast to previous results using the QMC model. Moreover,  those states have significant widths \cite{carmen10}, in particular, for $^{208}$Pb \cite{qmc}. Only $D^0$-nucleus bound states are possible since the Coulomb interaction prevents the formation of observable bound states for $D^+$ mesons. A similar study for $\bar{D}$ mesic atoms and nuclei is presented in Ref.~\cite{GarciaRecio:2011xt}.

 It is also interesting to note that a recent work suggests the possibility that, for the lightest nucleus, $DNN$ develops a bound and narrow state with $S=0, I=1/2$, as evaluated in Ref.~\cite{bayar}. 


\begin{figure}
\begin{center}
\includegraphics[width=0.5\textwidth]{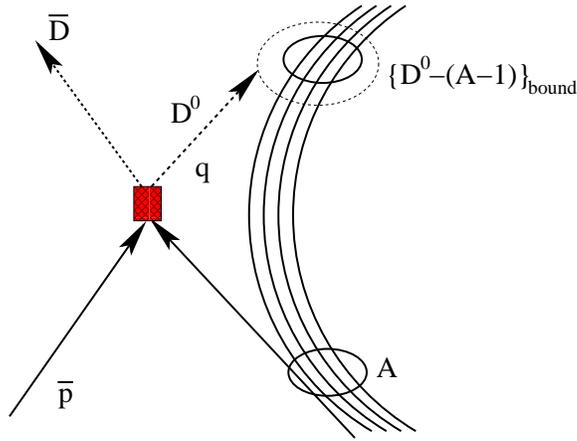}
\caption{Possible production of $D^0$-mesic nuclei with an antiproton beam\label{fig5}}
  \end{center}
\end{figure}


To gain some knowledge on the charm meson-nucleus interaction, the information on bound states is very valuable. This is of special interest  for PANDA at FAIR. The experimental detection of $D$ bound states is, however, a difficult task. For example, it was observed in Ref.~\cite{carmen10} that reactions with antiprotons on nuclei for obtaining $D^0$-nucleus states 
might have a very low production rate (see Fig.~\ref{fig5}). Reactions but with proton beams seem more likely to trap a $D^0$ in nuclei \cite{carmen10}. 

\section{Summary}

We have reviewed the properties of strange and charm mesons in dense matter. Different frameworks have been discussed paying a special attention to unitarized coupled-channel approaches that take, as bare interaction,  effective Lagrangians that respect chiral symmetry in the light sector while HQSS in the heavy one, and/or vector-meson exchange models. Moreover, we have analyzed possible experimental signatures of the in-medium properties of these mesons. In particular, we have presented heavy-ion collisions data on strange mesons, the transparency ratio of the $\gamma A \to K^+ K^{*-} A^\prime$ reaction and the study the possible formation of $D$-mesic nuclei.

\section*{Acknowledgments}

 This research was supported by DGI and FEDER
funds, under Contract Nos. FIS2011-28853-C02-02,
 FIS2011-24149, FIS2011-24154, FPA2010-16963 and the
Spanish Consolider-Ingenio 2010 Programme CPAN
(CSD2007-00042), by Junta de Andaluc\' ia Grant
No. FQM-225, by Generalitat Valenciana under Contract
No. PROMETEO/2009/0090, by the Generalitat de Catalunya under
contract 2009SGR-1289 and by the EU
HadronPhysics3 project, Grant Agreement No. 283286.
 L. T. acknowledges support
from Ramon y Cajal Research Programme, and from FP7-
PEOPLE-2011-CIG under Contract No. PCIG09-GA-
2011-291679.








\end{document}